\documentclass[11pt]{article}
\usepackage[dvips]{color}
\usepackage{epsfig}
\usepackage{amsmath}
\usepackage{graphicx}
\date{}
\begin{document}
\title{{\bf Signature transition in Einstein-Cartan cosmology}}
\author{Babak Vakili$^{1}$\thanks{b-vakili@iauc.ac.ir}\,\, and\,\,\
Shahram Jalalzadeh
 $^{2}$\thanks{s-jalalzadeh@sbu.ac.ir} \,\,\,\\\\$^1${\small {\it Department of Physics, Chalous Branch, Islamic Azad
University (IAU), P.O. Box 46615-397, Chalous, Iran }}\\$^2${\small
{\it Department of Physics,  Shahid Beheshti University, G. C.,
Evin, Tehran 19839,  Iran}} $^{}${}} \maketitle
\begin{abstract}
In the context of Einstein-Cartan theory of gravity, we consider a
Friedmann-Lemaitre-Robertson-Walker (FLRW) cosmological model with
Weyssenhoff perfect fluid. We focus attention on those classical
solutions that admit a degenerate metric in which the scale factor
has smooth behavior in the transition from a Euclidean to a
Lorentzian domain. It is shown that the spin-spin contact
interaction enables one to obtain such a signature changing
solutions  due to the
Riemann-Cartan ($U_4$) structure of space-time.\vspace{5mm}\noindent\vspace{0.8mm}\newline
\end{abstract}

\section{Introduction}
The choice of a matter field which is coupled with the Einstein
equations through its energy-momentum tensor has always a direct
effect on the study of the cosmological models. Traditionally, a
perfect fluid is usually used as the matter source. However, one can
not deny the constantly increasing role of the scalar fields in more
recent cosmological models as the matter source \cite{Scalar}. This
of course is expected since it is somewhat easier to work with
scalar field. It is also possible to imagine a Universe filled with
a classical spin fluid or even a massless or massive spinor fields
as the matter source. Such cosmological models have rarely been
studied in the literature and, when they were, it was more often
than not in the form of general formalisms \cite{spinor}. In general
then, it would be fair to say that cosmologies with spinor fields as
the matter source are the least studied scenarios. In 1923 \'{E}lie
Cartan introduced the relation between the intrinsic angular
momentum of matter and the space-time torsion in the framework of a
generalization of general relativity (GR) \cite{cartan1}, nowadays
known as Einstein-Cartan (EC) theory \cite{cartan}. Indeed, there
are two different methods to introduce the classical spin in GR. In
the first approach, spin is considered as a dynamical quantity
without changing the Riemannian structure of the space-time geometry
\cite{spin1}. The second method, which as we mentioned above was
proposed by Cartan, is based on the generalization of space-time
structure by assuming the metric and the non-symmetric affine
connection as independent quantities. Since the first attempts of
Cartan to bring spin into the curved space-time, many efforts have
been made in this area and the corresponding results have been
followed and developed by a number of works, see for instance
\cite{Hehl}, \cite{Trautman} and \cite{Kop}. The importance of the
Cartan theory becomes more clear, if one tries to incorporate the
spinor field into the torsion-free general theory of relativity. In
this context one should apply the Cartan theory which possesses
torsion as well as curvature \cite{Wat}. In EC theory, torsion is
not a dynamical quantity, instead it can be expressed completely in
terms of the spin sources \cite{Hehl}. Consequently, in order to
study the effects of torsion in $U_4$ geometry (it is usual to
denote the Riemann-Cartan space-time as $U_4$ to distinguish it from
the Riemannian space-time) one may consider the matter fields with
intrinsic angular momentum. To do this, one of the usual ways is to
consider a fluid with intrinsic spin density known as the
Weyssenhoff exotic perfect fluid \cite{Weys}. As in the case of
other alternative theories of gravity, it is important to seek the
cosmological solutions in the EC theory of gravity, i.e., in a
theory in which the spin properties of matter and their influence on
the geometrical structure of space-time are considered. This is done
by some authors \cite{Kuch}, who have investigated the effects of
torsion and spinning matter in a cosmological setting and its
possible role to remove the singularities, inflationary scenarios,
explain the late time accelerated expansion of the Universe and so
on.

An interesting topic related to classical and quantum cosmology is
that of signature change which has attracted attention since the
early 1980s. Traditionally, a feature in GR is that one usually
chooses a Lorentzian signature for the space-time metric before
attempting to solve the Einstein''s field equations. However, the
reason for doing so is not pre-determined and it is well known that
the field equations do not demand this property, that is, if one
ignores this requirement one may find solutions to the field
equations which, when parameterized suitably, can either have
Euclidean or Lorentzian signature. The notion of signature
transition first appeared in the works of Hartle and Hawking
\cite{Haw} where they argued that in quantum cosmology amplitudes
for gravity should be written as the sum of all compact Riemannian
geometries whose boundaries are located at the signature changing
hypersurface. Since then this subject has been studied at the
classical and quantum cosmology level by other authors, see for
example \cite{Sig}. In what follows by a signature changing
space-time we mean a manifold which contains both Euclidean and
Lorentzian region. As it is shown in \cite{Dray}, in classical GR, a
signature changing metric should be either degenerate or
discontinuous, though Einstein's equations implicitly assume that
the metric is non-degenerate and at least continuous.

In this letter, we consider a smooth signature changing type of flat
FLRW space-time in the framework of EC gravity with exotic
Weyssenhof perfect fluid. For the case of a spatially flat Universe,
field equations are then solved exactly for the scale factor as
dynamical variable, giving rise to cosmological solutions with a
degenerate metric, describing a continuous signature transition from
a Euclidean domain to a Lorentzian space-time.

\section{The model}
In this section we start by briefly studying the EC gravity where
the action is given by (we work in units where $c=1$ and consider
the signature $(+,-,-,-)$ for the space-time metric)
\begin{eqnarray}\label{1}
{\cal S}=\int\sqrt{-g}d^4x\left[-\frac{1}{16\pi G}\left(\tilde
R-2\Lambda\right)+{\cal L}_M\right],
\end{eqnarray}
where $\tilde R$ is the Ricci scalar constructed by the asymmetric
connection $\tilde\Gamma^\mu_{\,\,\,\,\alpha\beta}$ and $\Lambda$ is
the cosmological constant. By using of the metricity condition
\cite{Hehl}
\begin{equation}
\tilde\nabla_\alpha g_{\mu\nu}=0,\end{equation} and also the
definition of torsion
\begin{equation}
T^\mu_{\,\,\,\,\alpha\beta}:=\tilde\Gamma^\mu_{\,\,\,\,\alpha\beta}-\tilde\Gamma^\mu_{\,\,\,\,\beta\alpha},\end{equation}
the connection $\tilde\Gamma^\mu_{\,\,\,\,\alpha\beta}$ can be
expressed as
\begin{eqnarray}\label{2}
\tilde\Gamma^\mu_{\,\,\,\,\alpha\beta}=\Gamma^\mu_{\hspace{.2cm}\alpha\beta}+K^\mu_{\hspace{.2cm}\alpha\beta},
\end{eqnarray}
where $\Gamma^\mu_{\,\alpha\beta}$ is the Levi-Civita connection
(Christoffel symbol) and $K^\mu_{\hspace{.2cm}\alpha\beta}$ is the
contorsion tensor related to the torsion
$Q_{\alpha\beta}^{\hspace{.5cm}\mu}:=\tilde
\Gamma_{[\alpha\beta]}^{\hspace{.5cm}\mu}$ via
\begin{eqnarray}\label{3}
K^\mu_{\hspace{.2cm}\alpha\beta}:
=\frac{1}{2}\left(Q_{\hspace{.2cm}\alpha\beta}^{\mu}-Q^{\hspace{.3cm}\mu}_{\alpha\hspace{.4cm}\beta}-Q^{\hspace{.3cm}\mu}_{\beta\hspace{.4cm}\alpha}\right).
\end{eqnarray}
Also ${\cal L_M}$ is essentially the Lagrangian density for matter
field coupled to gravity. Our assumption is that instead of usual
Big-Bang singularity in the early Universe, we have signature
changing event. Therefore, we focus our attention on the early
Universe epoch where the matter content of the model is of the form
of fermionic matter, like quarks and leptons. The dynamical
equations of motion can be obtained by performing the variation of
the action with respect to the metric and contorsion \cite{Hehl},
that is

\begin{eqnarray}\label{4}
\left\{
\begin{array}{ll}
G^{\mu\nu}-\Lambda
g^{\mu\nu}-\left(\nabla_{\alpha}+2Q_{\alpha\beta}^{\hspace{.4cm}\beta}\right)\left(T^{\mu\nu\alpha}-T^{\nu\alpha\mu}+T^{\alpha\mu\nu}\right)=8\pi
GT^{\mu\nu},\\\\
T^{\mu\nu\alpha}=8\pi G\tau^{\mu\nu\alpha},
\end{array}
\right.
\end{eqnarray}
where
\begin{eqnarray}\label{5}
T_{\mu\nu}^{\hspace{.3cm}\alpha}=Q_{\mu\nu}^{\hspace{.3cm}\alpha}+\delta_\mu^\alpha
Q_{\nu\beta}^{\hspace{.4cm}\beta}-\delta_\nu^\alpha
Q_{\mu\beta}^{\hspace{.4cm}\beta},
\end{eqnarray}and $G^{\mu\nu}$ and $\nabla_\alpha$ are respectively the Einstein tensor and covariant derivative for the full
nonsymmetric connection $\Gamma$. Also
\begin{eqnarray}\label{6}
\left\{
\begin{array}{ll}
T^{\mu\nu}:=\frac{2}{\sqrt{-g}}\frac{\delta{\cal L}_M}{\delta g_{\mu\nu}},\\
\\
\tau^{\mu\nu\alpha}:=\frac{1}{\sqrt{-g}}\frac{\delta{\cal
L}_M}{\delta K_{\alpha\nu\mu}},
\end{array}
\right.
\end{eqnarray}
are the energy-momentum and the canonical spin-density tensors
respectively. Now by using equations (\ref{4}) and (\ref{5}) one can
obtain modified Einstein field equations
\begin{eqnarray}\label{7}
G^{\mu\nu}(\Gamma)=8\pi G(T^{\mu\nu}+\tau^{\mu\nu}),
\end{eqnarray}
where $G^{\mu\nu}(\Gamma)$ is the usual symmetric Einstein tensor
and
\begin{eqnarray}\label{8}
\tau^{\alpha\beta}=
\left[-4\tau^{\alpha\mu}_{\hspace{.4cm}[\nu}\tau^{\beta\nu}_
{\hspace{.3cm}\mu]}-2\tau^{\alpha\mu\nu}\tau^\beta_{\hspace{.3cm}\mu\nu}+\tau^{\mu\nu\alpha}\tau_{\mu\nu}^{\hspace{.3cm}\beta}+\frac{1}{2}g^{\alpha\beta}
\left(4\tau_{\lambda
\hspace{.1cm}[\nu}^{\hspace{.2cm}\mu}\tau^{\lambda\nu}_{\hspace{.4cm}\mu]}+\tau^{\mu\nu\lambda}\tau_{\mu\nu\lambda}\right)\right],
\end{eqnarray}
is the correction to the space-time curvature due to the spin
\cite{Weys}. If the spin vanishes then equation (\ref{7}) reduces to
the standard Einstein field equations. We assume that ${\cal L}_M$
describes a fluid of spinning particles in the early Universe
minimally coupled to the metric and the torsion of the $U_4$ theory.
For the spin fluid the canonical spin tensor is given by \cite{Weys}
\begin{eqnarray}\label{9}
\tau^{\mu\nu\alpha}=\frac{1}{2}S^{\mu\nu}u^\alpha,
\end{eqnarray}
where $S^{\mu\nu}$ is the antisymmetric spin density and $u^\alpha$
is the 4-velocity of the fluid \cite{Ob}. Then the energy-momentum
tensor can be decomposed into the two parts: the usual perfect fluid
$T_F^{\hspace{.1cm}\alpha\beta}$ and an intrinsic-spin part
$T_S^{\hspace{.1cm}\alpha\beta}$, as
\begin{equation}T^{\alpha\beta}
= T_F^{\hspace{.1cm}\alpha\beta}+
T_S^{\hspace{.1cm}\alpha\beta},\end{equation} so that we have
explicitly for intrinsic-spin part
\begin{eqnarray}\label{10}
T_S^{\hspace{.2cm}\alpha\beta}=u^{(\alpha}S^{\beta)\mu}u^\nu
u_{\mu;\nu}+(u^{(\alpha}S^{\beta)\mu})_{;\mu}+Q_{\mu\nu}^{\hspace{.2cm}(\alpha}u^{\beta)}
S^{\nu\mu}-u^\nu
S^{\mu(\beta}Q^{\alpha)}_{\hspace{.1cm}\mu\nu}-\omega^{\mu(\alpha}S^{\beta)}_{\hspace{.3cm}\mu}+u^{(\alpha}S^{\beta)\mu}\omega_{\mu\nu}u^\nu,
\end{eqnarray}
where $\omega$ is the angular velocity associated with the intrinsic
spin and semicolon denotes covariant derivative with respect to
Levi-Civita connection. If as usual interpretation of EC gravity we
assume that $S_{\mu\nu}$ is associated with the quantum mechanical
spin of microscopic particles \cite{Kuch}, then for unpolarized
spinning field we have $<S_{\mu\nu}>=0$ and if we define
\begin{equation}
\sigma^2:=\frac{1}{2}<S_{\mu\nu}S^{\mu\nu}>,\end{equation} we get
\begin{eqnarray}\label{11}
<\tau^{\alpha\beta}>=4\pi G \sigma^2u^\alpha u^\beta +2\pi G
\sigma^2g^{\alpha\beta},
\end{eqnarray}
and
\begin{eqnarray}\label{12}
\left\{
\begin{array}{ll}
<T_F^{\hspace{.1cm}\alpha\beta}>=(\rho+p)u^\alpha u^\beta-pg^{\alpha\beta},\\\\
<T_S^{\hspace{.2cm}\alpha\beta}>=-8\pi G\sigma^2u^\alpha u^\beta.
\end{array}
\right.
\end{eqnarray}
Consequently the simplest EC generalization of standard gravity will
be
\begin{eqnarray}\label{13}
G^{\alpha\beta}(\Gamma)=8\pi G \Theta^{\alpha\beta},
\end{eqnarray}
where $\Theta^{\alpha\beta}$ describes the effective macroscopic
limit of matter field
\begin{eqnarray}\label{14}
\Theta^{\alpha\beta}:=<T^{\alpha\beta}>+<\tau^{\alpha\beta}>=\left(\rho+p-4\pi
G\sigma^2\right)u^\alpha u^\beta-\left(p-2\pi
G\sigma^2\right)g^{\alpha\beta}.
\end{eqnarray}In analogy with the usual GR, equations
(\ref{13}) and (\ref{14}) show that EC field equations are
equivalent to the Einstein equations coupled to a fluid with a
particular equation of state as the matter source. Indeed, in a
hydrodynamical description the contribution of the torsion can be
carried out by a spin fluid such that
\begin{equation}\label{rev1}
\rho_{tot}=\rho-2\pi G \sigma^2,\hspace{5mm}p_{tot}=p-2\pi G
\sigma^2.
\end{equation} It is important to note that the signs of the correction
terms in (\ref{rev1}) are negative which is in agreement with the
semi-classical models of spin fluid \cite{Weys}, \cite{Kuch}. This
means that the effect of spin in EC theory is like a perfect fluid
with negative energy density and pressure. In what follows, we will
see that these negative signs are required to get the signature
changing solutions. However, we would like to emphasize that our
model, in some senses, is different with the model considered in
\cite{Wat} in which a Dirac field plays the role of a spin fluid
with positive energy density. In such a model, although under some
conditions an accelerated expansion of the universe will occur, the
metric of space-time does not experience a change of signature and
hence the problem of the initial singularity is still not resolved.
\section{Signature changing cosmology}
\begin{figure}
\centering
\includegraphics[width=5cm]{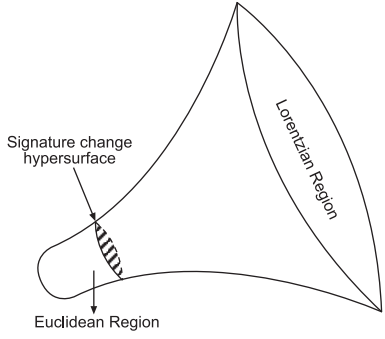}
\caption{A typical signature changing manifold.}\label{fig1}
\end{figure}

According to the Hartle-Hawking no-boundary proposal \cite{Haw}
space-time is partly Euclidean and partly Lorentzian (see figure
\ref{fig1}). The main motivation for this idea is the path integral
formulation of quantum gravity. To have a better understanding of
the quantum theory it is necessary to have an understanding of the
associated classical theory by constructing the classical space-time
with signature changing structure. In fact, there are two main
proposals for this purpose. In the first proposal, the metric of
space-time is everywhere non-degenerate but fails to be continuous
at the signature changing hypersurface that divides the Euclidean
from the Lorentzian region. On the other hand, in the second
proposition, the metric is smooth everywhere but is degenerate at
the hypersurface of signature change \cite{Ell}. Here, we are
interested in using the second one. The authors of \cite{Koss} have
shown that for smooth signature changing space-time there exist
coordinates such that
\begin{eqnarray}\label{15}
ds^2=tdt^2-h_{ij}dx^idx^j.
\end{eqnarray}

For this case, Kossowski and Kriele \cite{Koss1} have shown that the
in GR energy-momentum tensor of the matter field becomes bounded if
and only if the signature change hypersurface $(\Sigma)$ is totally
geodesic and $\partial_t h_{ij}=0$ at $\Sigma$. To proceed further,
let us consider the signature changing FLRW metric as
\begin{eqnarray}\label{16}
ds^2=tdt^2-a(t)^2g_{ij}dx^idx^j,
\end{eqnarray}
where $g_{ij}$ is the metric on the constant-curvature spatial
section. Inserting this signature changing line element into the
(\ref{13}) and (\ref{14}) gives the field equations
\begin{eqnarray}\label{17}
\left\{
\begin{array}{ll}
\frac{1}{t}\left(\frac{\dot{a}}{a}\right)^2+\frac{k}{a^2}=\frac{8\pi
G}{3}\rho-\frac{(4\pi
G)^2}{3}\sigma^2, \\\\
\frac{1}{t}\frac{\ddot a}{a}-\frac{1}{2t^2}\frac{\dot
a}{a}=-\frac{4\pi G}{3}(3p+\rho)+\frac{2}{3}(4\pi
G)^2\sigma^2, \\
\end{array}
\right.
\end{eqnarray}
where dot denotes the derivation with respect to  $t$ and $k$
defines the curvature of the spatial section, taking the values $0$,
$1$, $-1$ for a flat, positive-curvature or negative-curvature
Universe, respectively. The combination of field equations
(\ref{17}) gives
\begin{eqnarray}\label{18}
\frac{d}{dt}(\rho-2\pi G\sigma^2)=-3\frac{\dot a}{a}(\rho+p-4\pi
G\sigma^2),
\end{eqnarray}
which is a generalization of the covariant energy conservation law
to include the spin. Now, we consider the matter field as a
unpolarized fermionic perfect fluid with equation of state
$p=\gamma\rho$. Consequently, we have
\begin{equation}
\sigma^2=\frac{1}{2}<S^2>=\frac{1}{8}\hbar^2<n^2>,\end{equation}
where $n$ denotes the particle number density, and averaging
procedure gives \cite{Nurgaliev}
\begin{eqnarray}\label{19}
\sigma^2=\frac{\hbar^2}{8}B_\gamma^{-\frac{2}{1+\gamma}}\rho^{\frac{2}{1+\gamma}},
\end{eqnarray}
where $B_\gamma$ is a dimensional constant dependent on $\gamma$.
Therefore, conservation equation (\ref{18}) gives
\begin{eqnarray}\label{20}
\rho=\rho_0a^{-3(1+\gamma)},
\end{eqnarray}
where $\rho_0$ is energy density at present time. If we define for
simplicity
\begin{equation}
C:=\frac{4\pi G}{3}\rho_0,\hspace{5mm} D:=\frac{(4\pi G)^2}{24}\hbar
B_\gamma^{-\frac{2}{1+\gamma}}\rho_0^{\frac{2}{1+\gamma}},\end{equation}
then the Friedmann equation (\ref{17}) will be
\begin{eqnarray}\label{21}
\left\{
\begin{array}{ll}
\frac{1}{t}\left(\frac{\dot{a}}{a}\right)^2+\frac{k}{a^2}=2Ca^{-3(1+\gamma)}-Da^{-6},\\\\
\frac{1}{t}\frac{\ddot{a}}{a}-\frac{1}{2t^2}\frac{\dot{a}}{a}=-C(1+3\gamma)a^{-3(1+\gamma)}+2Da^{-6}.
\end{array}
\right.
\end{eqnarray}
The avoidance of the singularity is due to the repulsive force
$F:=-\partial_a(-D/a^3)$ extracted from the spinning matter
potential. In fact the quantum mechanical nature of spin induces the
negative pressure which is important at the very early Universe and
is responsible for the existence of signature change hypersurface.
From now on we will focus our attention to the special case $k=0$,
for which the sign of the left-hand side of the first Friedmann
equation is negative for the negative values of $t$ and positive for
$t>0$. Consequently, the sign of the right-hand side changes as
well. Hence the right-hand side vanishes at $t=0$. Now, since we
have solutions both for $t<0$ and $t>0$, there should therefore
exist signature changing hypersurface so that
$a=[\frac{D}{2C}]^{\frac{1}{3(1-\gamma)}} =a_0$. Also, it is easy to
see from Friedmann equation that the scale factor is less than $a_0$
for negative values of $t$ and grater than $a_0$ when $t$ is
positive. Hence, this equation predicts the existence of three
regions, namely, a Lorentzian domain, a signature changing
hypersurface, and an Euclidean domain.

\begin{figure}
\centering
\includegraphics[width=3.5cm]{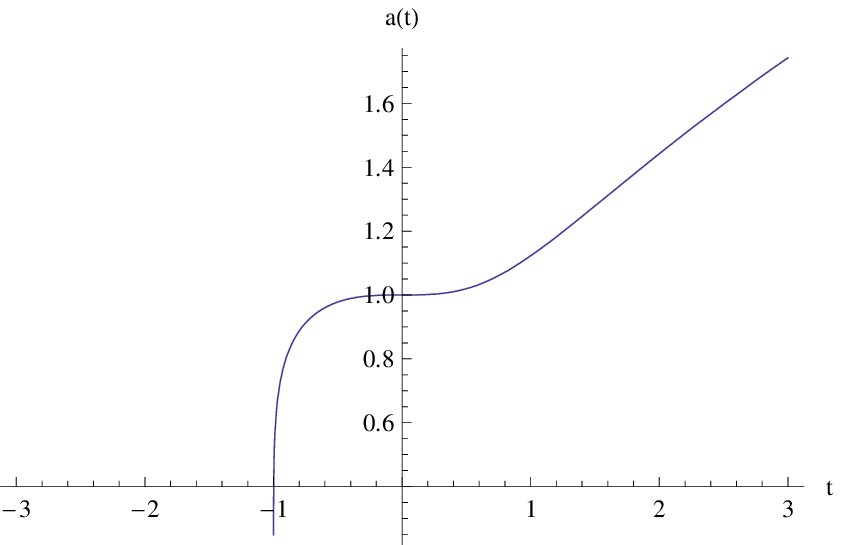}\hspace{2mm}
\includegraphics[width=3.5cm]{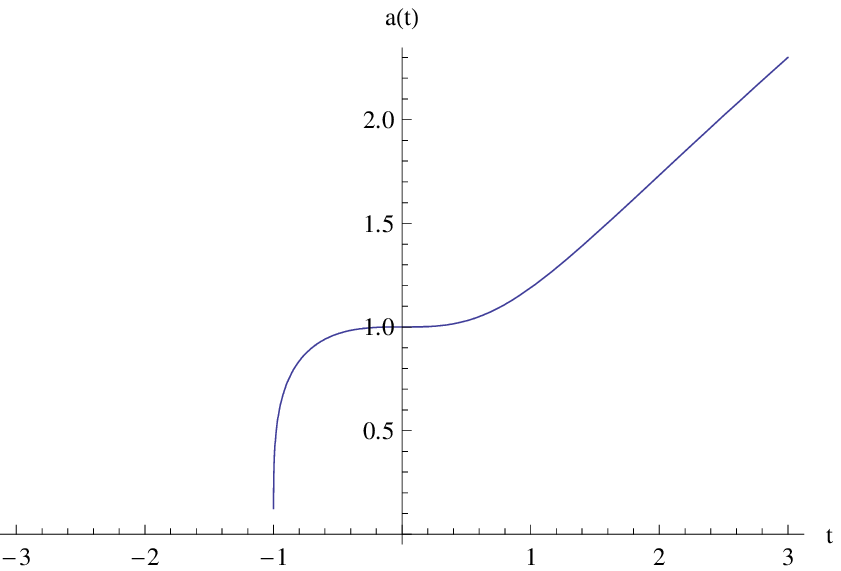} \hspace{2mm} \includegraphics[width=3.5cm]{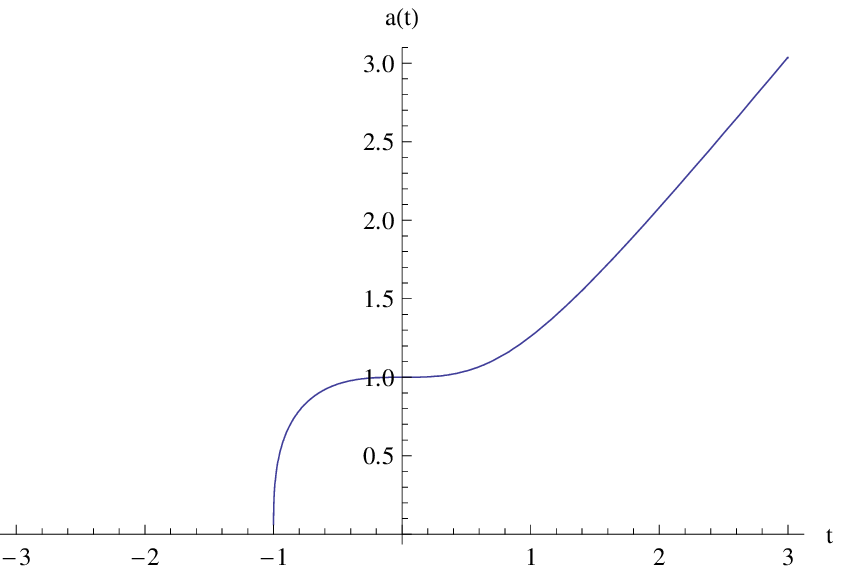}\hspace{2mm}
\includegraphics[width=3.5cm]{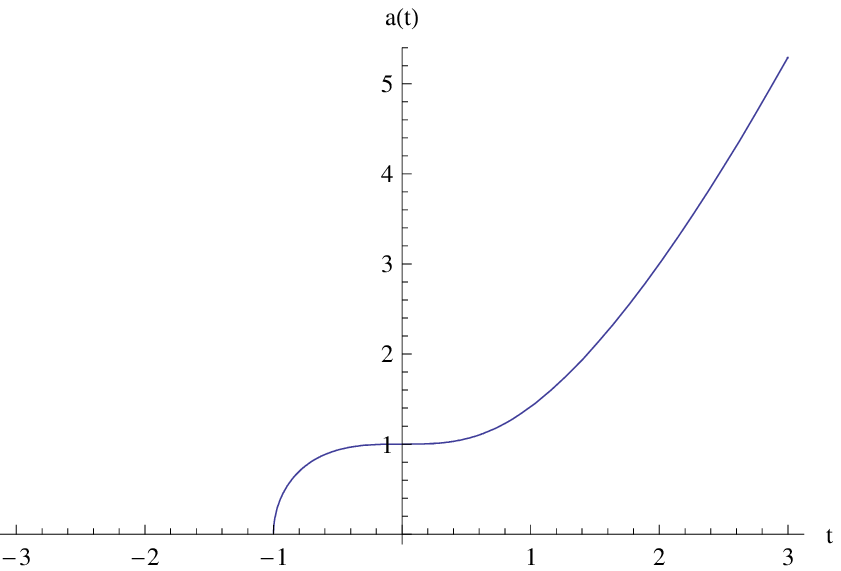}
\caption{Qualitative behavior of the scale factor versus time based
on relation (\ref{27}) for typical numerical values of the
parameters. The figures are plotted for $\gamma=-1,-1/3,0,1/3$ from
left to right.}\label{fig2}
\end{figure}

The exact solution of the Friedmann equations (\ref{21}) for a flat
$(k=0)$ distribution of dust $(\gamma=0)$ reads
\begin{eqnarray}\label{23}
a(t)=\left(\frac{D}{2C}\right)^{\frac{1}{3}}\left[1+\frac{8C^2}{D}t^3\right]^{\frac{1}{3}},
\end{eqnarray}
which shows a continuous transition from a finite Euclidean domain
to the Lorentzian one. Another exact solution in flat case is
radiation dominated $U_4$ Universe
\begin{eqnarray}
a\sqrt{a^2-\frac{D}{2C}}+\frac{D}{2C}\ln\left[{a+\sqrt{a^2-\frac{D}{2C}}}\right]=\sqrt{2C}\frac{4}{3}t^{\frac{3}{2}}.
\end{eqnarray}
It is clear that in the radiation case one cannot explicitly write
the scale factor in terms of $t$. To obtain solution of Friedmann
equations close to the signature changing hypersurface, we can use
signature changing conformal time
\begin{eqnarray}\label{25}
\sqrt{t}dt=a^{3\gamma}\sqrt{\eta}d\eta,
\end{eqnarray}
which leads to the following solution\begin{eqnarray}
a(\eta)=\left(\frac{D}{2C}\right)^{-\frac{1}{3(\gamma-1)}}\left[\frac{4C^2}{81(\gamma-1)^2D}\eta^3+1\right]^{-\frac{1}{3(\gamma-1)}}.
\end{eqnarray}
To write the scale factor in terms of signature changing time $t$,
one may expand the above solution around the signature changing
hypersurface $\Sigma$ which upon integration the relation
(\ref{25}), that is,
\begin{eqnarray}\label{26}
\eta^3=\left(\frac{D}{2C}\right)^{\frac{2\gamma}{\gamma-1}}t^3,
\end{eqnarray}
results the following expression for the scale factor close to the
signature changing hypersurface
\begin{eqnarray}\label{27}
a(t)=\left(\frac{D}{2C}\right)^{-\frac{1}{3(\gamma-1)}}
\left[1+\frac{(2C)^{\frac{2}{1-\gamma}}}{81(\gamma-1)^2D^{\frac{1+\gamma}{1-\gamma}}}t^3\right]^{-\frac{1}{3(\gamma-1)}}.
\end{eqnarray}
Also, it is easy also to see that
\begin{eqnarray}\label{28}
\frac{\partial a(t)}{\partial t}|_{t=0}=0,
\end{eqnarray}
\begin{eqnarray}\label{29}
\frac{\partial^2 a(t)}{\partial t^2}|_{t=0}=0,
\end{eqnarray}
which satisfy the Kossowski and Kriele theorem mentioned above. As
we have shown in figure \ref{fig2}, the above solution like
(\ref{23}) shows a continuous transition from a finite Euclidean
domain to a Lorentzian one.

\section{Summary}
In this letter, we have shown that the EC cosmological model
predicts a signature change when the singularity is approached.
Moreover, the spinning matter leads to a repulsive force which
results in a regular transition from Euclidean to the Lorentzian
region. The above discussion shows that one of the curious features
of quantum cosmology is the use of Riemannian signature spaces to
explain the origin of the observable Lorentzian signature Universe.
There are various interpretations of this, the simplest of which is
that the signature of the Universe was initially Riemannian and then
subsequently changed. It may be argued that the Lorentzian signature
is an independent assumption of relativity rather than a
consequence, with the theory being equally valid for Riemannian
signature, and that in a quantum theory of gravity it would be
unnatural to impose signature restrictions on the metric. The
question arises as to whether the qualitative predictions of quantum
cosmology can be obtained from purely classical relativity by
relaxing the assumption of Lorentzian signature. Also in order to
understand the quantum theory it is necessary to have an
understanding of the associated classical theory, i.e., the theory
of classical space-times with signature type change.

Finally we want to point out in connection with the EC theory of
gravity and the tetrad (vierbein) formalism which is required for
the coupling of spin to gravity. In view of the construction of the
field equations the tetrad and the spin connection are considered as
independent fields in the action of the theory. As is well known in
terms of a tetrad orthonormal frame $e^{\mu}_a(x)$ the space-time
metric at any point can take the form of the Minkowski metric.
Hence, in the tetrad formalism the metric signature seems to be
fixed in the signature of the Minkowski metric. Now, a question
arises: Is it possible to consider the issue of signature transition
in this formalism? To answer this question, note that while the
space-time metric has 10 components, the tetrad field has 16
components. Indeed, renunciation of the strong equivalence principle
in favor of the Galilei-E\"{o}tv\"{o}s principle makes it possible
to introduce in gravitational theory more field components than the
ten independent components of the space-time metric. In the tetrad
formalism of GR all of the 16 components $e^{\mu}_a(x)$ are employed
to serve as gravitational field potentials and the effect of
gravitation on the matter is represented in a locally
Lorentz-covariant manner. Therefore, the tetrad formalism of EC
theory, as respect to the fixed signature of Minkowski space-time,
is not suitable to survey signature change phenomena.
\vspace{5mm}\newline \noindent {\bf
Acknowledgement}\vspace{2mm}\noindent\newline The authors are
grateful to the research council of IAU, Chalous Branch for
financial support.

\end{document}